\documentclass[conference]{IEEEtran}
\newtheorem{definition}{Definition}

\ifCLASSINFOpdf
 \usepackage[pdftex]{graphicx}
\else
\fi

\begin{document}
\title{A Modeling Framework for Generating Security Protocol Specifications}


\author{\IEEEauthorblockN{Genge B\'{e}la and Haller Piroska}
\IEEEauthorblockA{Electrical Engineering Department\\
``Petru Maior'' University of T\^{a}rgu Mure\c{s}\\
Nicolae Iorga str., No. 1, 540088, Mure\c{s}, ROMANIA\\
\{bgenge,phaller\}@upm.ro}
}


%Telephone: (800) 555--1212, Fax: (888) 555--1212}


\maketitle

\begin{abstract}
%\boldmath
We propose a modeling framework for generating security
protocol specifications. The generated protocol specifications rely on the use of a sequential and a semantical component. The first component defines protocol properties such as preconditions, effects, message sequences and it is developed as a WSDL-S specification. The second component defines the semantic aspects corresponding to the messages included in the first component by the use of ontological constructions and it is developed as an OWL-based specification. Our approach was validated on 13 protocols from which we mention: the ISO9798 protocol, the CCITTX.509 data transfer protocol and the Kerberos symmetric key protocol.\newline


\end{abstract}



% \fi
\IEEEpeerreviewmaketitle

%-------------------------------------------------------------------------
\section{Introduction}

Security protocols are widely used today to provide secure
communication over insecure environments. By examining the
literature we come upon various security protocols designed to provide solutions to specific problems \cite{WWW:SPORE}. With this large amount of protocols to chose from, distributed heterogenous systems must be prepared to handle multiple security protocols.

Existing technologies, such as the Security Assertions
Markup Language \cite{WWW:SAML} (i.e. SAML) or WS-Security \cite{WWW:WSSECURITY} provide a unifying solution for the authentication and authorization issues through the use of
predefined protocols. By implementing these protocols, Web services authenticate users and provide authorized access to resources. However, despite the fact that existing solutions provide a way to implement security claims, these approaches
are rather static. This means that in case of new security protocols, services must be reprogrammed.

In this paper we propose a more flexible solution to this problem by developing a security protocol specification generation framework based on existing Web service technologies such as WSDL-S \cite{WWW:WSDLS} and OWL \cite{WWW:OWL}, aiming at the automatic discovery and execution of security protocols.

A security protocol specification is a description of the protocol messages exchanged by participants and of the mechanisms related to the construction and processing
of messages. By inspecting the literature we come upon various forms of specifications \cite{ART:INFSPEC,ART:SPI}, each specification being specifically designed for a task.

Based on this observation, in the process of developing a new
specification we first formulate a set of requirements. Because the proposed specification includes a description of the messages exchanged by participants, which is also one of the goals of the well-known \emph{informal} specification, we consider the informal specification as the starting point of the construction process. Based on the formulated requirements, we identify two components: the message sequential specification, or more briefly SEQ-S, and the semantic specification, or more briefly SEM-S.

The first component is designed as a WSDL-S  specification which includes the sequence of messages that must be executed. For each message component an annotation is provided in order to link the component with the corresponding semantic information.

The second component is designed as an ontological specification by using OWL. An ontology is a ``formal, explicit specification of a shared conceptualization'' \cite{ART:ONTOLOGY}, consisting of concepts, properties (i.e. relations) and restrictions. Ontologies are part of the semantic Web technology, which associates semantic descriptions to Web services. Each message component from the protocol is represented as a concept in a hierarchical structure. In order to provide processing information, domain-range properties are defined for each concept.

The rest of the paper is structured as follows. In section \ref{SEC:FRAMEWORK} we provide a description of the proposed framework. In section \ref{SEC:EXPRESULTS} we present some of our experimental results. In section \ref{SEC:RELWORK} we relate our work to others. We end with a conclusion and future work in section \ref{SEC:CONCLUSIONS}.

%-------------------------------------------------------------------------
\section{The proposed framework}\label{SEC:FRAMEWORK}

%------------------------------------------------------------------------
\subsection{Requirements}

The proposed framework must guarantee the construction of a complete security protocol specification. The basic requirements that must be satisfied are extracted from the \emph{informal specification}. These include the explicit specification of protocol participants, message directions, cryptographic algorithm classes and message component types.

In order for participants to implement and execute protocols based on the generated specifications, the above-mentioned requirements are not enough. We also need to include several requirements for describing internal mechanisms such as constructing, processing and verifying messages. The resulting additional requirements include, among others, specifying the cryptographic parameters and processing operations for constructed and processed message components.

Based on these requirements we identified two specification
components: the message sequence specification, or SEQ-S, and the semantic specification, or SEM-S.

%------------------------------------------------------------------------
\subsection{SEQ-S structure}

The message sequence specification has been developed as a WSDL-S specification. The WSDL-S specification inherits the structure of WSDL. It provides, in addition to WSDL, semantic annotation possibilities through the use of the \emph{wssem:modelReference} attribute. In SEQ-S, the WSDL-S sections are maintained and used without any change.

For each protocol participant a WSDL-S specification is constructed. Because of this, each WSDL-S specification will contain only one \emph{portType} section and one \emph{binding} section. The \emph{portType} section provides an abstract group of \emph{operations}. Each \emph{operation} contains a reference to a \emph{message} and an attribute containing the communication direction of that message (i.e. input or output).

In addition to providing semantics to each message, we use the
\emph{wssem:precondition} and \emph{wssem:effect} tags to specify semantic information related to the preconditions needed to be satisfied in order to execute the protocol and effects that are activated if the protocol is executed.

%------------------------------------------------------------------------
\subsection{SEM-S structure}

Preconditions, effects and XML schema elements are annotated in SEQ-S with references to concepts from the semantic specification (SEM-S). SEM-S is constructed as an ontology consisting of several smaller ontologies. In order to satisfy the requirements formulated in the previous sections we identified 7 ontologies based on which the semantic specification is constructed.

\begin{figure}[htb]
\begin{center}
\scalebox{0.50}{\includegraphics{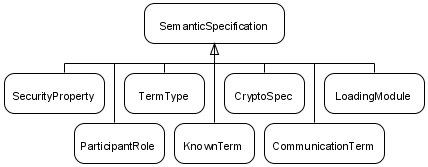}}
\caption{Core ontology of SEM-S}\label{FIG:SECPROTO}
\end{center}
\end{figure}

In the design of these ontologies we followed the principles
proposed in \cite{ART:ONTOLOGY}. As in any design process, we used a repetitive design and implementation in order to model the requirements of as many protocols as possible. The resulting ontologies are the starting point for constructing the semantic specification of a security protocol. Figure \ref{FIG:SECPROTO} shows the core ontology of SEM-S.

In the process of constructing SEM-S, the proposed sub-ontologies are extended with concepts and properties. The concepts are specific to each protocol, however, the defined properties are applied on all constructions. From these properties we mention: \emph{isOfType}, \emph{isEncrypted}, \emph{isStored}, \emph{isVerified}, \emph{isExtracted}, \emph{hasSymmetricAlgorithm} or \emph{hasKey}.

In the remaining of this sub-section we construct a formal ontology model used in the definition process of the proposed rules. The ontology model is defined as follows.\newline

\begin{definition}
\emph{An \emph{ontology model (OM)} is a triplet $\langle CONC,$ $PROP,INST\rangle$, where $CONC$ is the set of concepts defined for the ontology, $PROP$ is the set of properties and $INST$ is the set of all instances. An element from $PROP$ is a pair $\langle\alpha,\beta\rangle$, where $\alpha$ is a unique id and $\beta$ is a syntactic construction denoting the property name.}\newline
\end{definition}

In order to handle elements from an OM, we define the following mapping functions:
\begin{itemize}
  \item $domain:PROP\rightarrow CONC$ to map the domain concept corresponding to a given property;
  \item $range:PROP\rightarrow CONC$ to map the range concept corresponding to a given property;
  \item $prop:CONC\rightarrow PROP^*$ to map the set of properties for which the given concept is a domain;
  \item $parent:CONC\rightarrow CONC$ to map the parent of a given concept;
  \item $subcon:CONC\rightarrow CONC^*$ to map the set of concepts for which the given concept is a parent concept;
  \item $mincard:PROP\rightarrow \textsf{N}$ to map the minimum cardinality for a given property;
  \item $maxcard:PROP\rightarrow \textsf{N}$ to map the maximum cardinality for a given property.
\end{itemize}

The construction of SEM-S is based on a set of 13 rules we
identified using a repetitive design and implementation of
specifications. Because of space considerations in this section we only present rules that are the most relevant

\emph{Processing rules}. Rules from this category provide a set of guiding lines to model terms and protocol operations as concepts and properties such that processing of terms is made possible.

For example, the next rule states that for every concept from the \emph{KnownTerm} sub-ontology there must be an \emph{isOfType} property defined.\newline

\emph{$\mathbf{Rule 1}$. For every sub-concept of $KnownTerm$ that
is not a sub-concept of a cryptographic concept there is an
$isOfType$ property defined. Formally,}

$\forall c \in subcon(KnownTerm)\, if$

$\not\exists p'\in prop(parent(c)):$

$name(p)=SymmEncrypted\,, then\,\exists p \in prop(c):$

$\quad name(p)=isOfType \wedge$

$\quad mincard(p)=maxcard(p)=1$.\newline

\emph{Cryptographic rules}. The rules from this category require that for each generated term (e.g. symmetric key, random number) or cryptographic term (e.g. symmetric encryption, hash, signature), generation or construction properties are also specified.\newline

\emph{$\mathbf{Rule 2}$. For every sub-concept of $GeneratedTerm$
with the $RandomNumber$ property defined, there is a $hasLength$
property. Formally,}

$\forall c \in subcon(GeneratedTerm)\,\, if\,\, \exists p \in prop(c):$

$name(p)=RandomNumber\,\, then\,\, \exists p' \in prop(c):$

$\quad name(p')=hasLength \wedge$

$\quad mincard(p')=maxcard(p')=1$.\newline

\emph{Storage rules}. In order to model storage modules from which keys, certificates or tokens are extracted we use the
\emph{LoadingModule} sub-ontology.\newline

\emph{$\mathbf{Rule3}$. For every sub-concept of $LoadedTerm$ there is an $isLoaded$ property defined. Formally,}

$\forall c \in subcon(LoadedTerm), \exists p \in prop(c):$

$\quad name(p)=isLoaded \wedge$

$\quad mincard(p)=maxcard(p)=1$.\newline

%------------------------------------------------------------------------
\subsection{Example construction}

In order to provide an example usage of the previously defined
framework we provide a partial construction of the sequential and semantic specification for the initiator role from the ``BAN concrete Andrew Secure RPC''.

By examining the informal specification we conclude that participant $A$ is the initiator of the protocol. For this participant, we must define two outgoing and two incoming messages. The goal of the protocol is the exchange of the session key $K$. The resulting partial SEQ-S specification is given in figure
\ref{FIG:WSDLPrecondEffect}.
\begin{figure}[h]
\begin{scriptsize}
\begin{verbatim}
...
 <xsd:element name="Msg1Request">
   <xsd:complexType>
     <xsd:sequence>
       <xsd:element name="Participant A" type="xsd:string"
         wssem:modelReference=".../SecProt.owl#Sent_A"/>
       <xsd:element name="Random" type="xsd:base64Binary"
         wssem:modelReference=".../SecProt.owl#Sent_Na/>
     </xsd:sequence>
   </xsd:complexType>
 </xsd:element>
 <xsd:element name="Msg2Response">
   <xsd:complexType>
     <xsd:sequence>
       <xsd:element name="EncTerm1" type="xsd:base64Binary"
         wssem:modelReference=".../SecProt.owl#EncTerm1">
       </xsd:element>
     </xsd:sequence>
   </xsd:complexType>
 </xsd:element>
 ...
 <wsdl:operation name="Msg1">
   <wsdl:output message="tns:Msg1Request"/>
 </wsdl:operation>
 <wsdl:operation name="Msg2">
   <wsdl:input message="tns:Msg2Response"/>
 </wsdl:operation>
 <wssem:effect name="SessionKeyExchange"
   wssem:modelReference=".../SecProt.owl#SessionKey"/>
 ...
\end{verbatim}
\end{scriptsize}
    \caption{SEQ-S example schema, precondition and effect}\label{FIG:WSDLPrecondEffect}
\end{figure}
\begin{figure}[htb]
\begin{center}
\scalebox{0.50}{\includegraphics{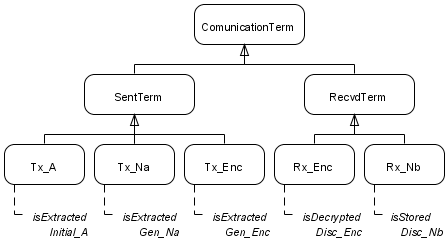}}
\caption{CommunicationTerm sub-ontology}\label{FIG:COMMTERMS}
\end{center}
\end{figure}
\begin{figure}[htb]
\begin{center}
\scalebox{0.50}{\includegraphics{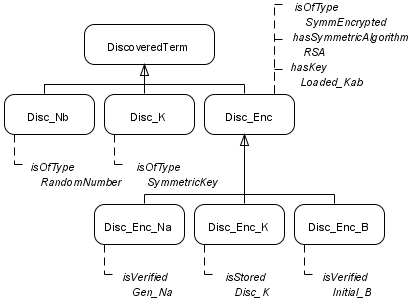}}
\caption{Discovered terms modeled as concepts}\label{FIG:DISCOVTERMS}
\end{center}
\end{figure}

The SEM-S construction process starts by examining the SEQ-S
specification from which the concepts that must be added to the \emph{CommunicationTerm} sub-ontology are extracted. The resulting \emph{CommunicationTerm} and \emph{KnownTerm} sub-ontologies are given in figure \ref{FIG:COMMTERMS} and figure \ref{FIG:DISCOVTERMS} respectively, where interrupted lines denote properties.

%------------------------------------------------------------------------
\section{Experimental results}\label{SEC:EXPRESULTS}

For our experiments, we developed over 38 WSDL-S and 38 OWL
specifications corresponding to initiator and respondent protocol roles. The WSDL-S specifications were constructed using Eclipse's web service package \cite{WWW:ECLIPSE} while the OWL specifications were constructed using the well-known ontology editor, Prot\'{e}g\'{e} \cite{ART:PROTEGE}.

The goal of our experiments was to prove that the specifications contain sufficient information for participants to execute the described protocols and at the end of the execution process the protocol goals are achieved.

In order to execute the specifications, messages were encoded and transmitted according to the constructions provided by the
WS-Security standard \cite{WWW:WSSECURITY}. In the experiments we conducted, participants downloaded the specification files from a public server and they were able to execute the protocols based only on the received descriptions. The participants hardware and software configurations: Intel Dual Core CPU at 1.8GHz, 1GByte of RAM, MS Windows XP.
\begin{table}
\centering \caption{Protocol execution timings}
\label{TAB:EXPRESULTS}       % Give a unique label
\begin{tabular}{|l|c|c|c|c|}
\hline
\textbf{Protocol}    &  \textbf{S-PR.} & \textbf{M-CON.} & \textbf{M-PR.}    & \textbf{Total} \\
\textbf{participant} & \textbf{(ms)}& \textbf{(ms)}& \textbf{(ms)} & \textbf{(ms)} \\
\hline\hline

BAN\,Init.    &  14.58 & 11.81 & 3.68 & 30.08 \\
BAN\,Resp.    &  14.03 & 2.86 & 1.62 & 18.52 \\
\hline

ISO9798\,Init.    &  13.07 & 35.784 & 23.30  & 72.16 \\
ISO9798\,Resp.    &  13.51 & 6.876 & 12.24 & 32.63 \\
\hline

Kerb.\,Init.\,1    &  22.63 & 0.83 & 0  & 23.47 \\
Kerb.\,Init.\,2    &  12.61 & 0.55 & 1.58& 14.76 \\
Kerb.\,Init.\,3    &  2.23 & 3.34 & 0.94& 6.52 \\
Kerb.\,Resp.\,1    &  19.28 & 0 & 0.41 & 19.69 \\
Kerb.\,Resp.\,2    &  10.81 & 3.379 & 1.67 & 15.87 \\
Kerb.\,Resp.\,3    &  5.25 & 11.41 & 3.59  & 20.26 \\
\hline
\end{tabular}
\end{table}

Part of the experimental results are given in table
\ref{TAB:EXPRESULTS}, where the values correspond to milliseconds. The S-PR column denotes the specification processing time, the M-CON column denotes the message construction time (for output messages) and the M-PR column denotes the message processing time (for input messages). The table contains two two-party protocols (``BAN concrete Andrew Secure RPC'', or more simply BAN, and ISO9798) and one three-party protocol (Kerberos). The performance differences
between the BAN and ISO9798 protocols are due to the fact that
ISO9798 makes use of public key cryptography, while BAN uses only symmetric cryptography.

%------------------------------------------------------------------------
\section{Related work}\label{SEC:RELWORK}

In this section we describe approaches we found in the literature that mostly relate to our proposal.

An approach that aims at the automatic implementation of security protocols is given in \cite{ART:AUTIMPLFORMAL}. This approach uses a formal description as a specification which is executed by participants. The proposed specification does not make use of Web service technologies, because of which inter-operability of systems executing the given specifications becomes a real issue. In addition, because our approach uses the ontology model, it benefits of several properties specific to ontologies, such as semantic properties, extendability or reusability of ontologies developed by others.

Another automated security protocol implementation approach is proposed in \cite{ART:AUTSPECIMPLFORMAL}. The specification in this case is constructed as an XML document from which the code is automatically generated. The resulting code is then compiled and executed by participants. Because of this aspect, our proposal is more dynamic in the sense that applications can download and execute new protocols based on the developed specifications automatically, without having to stop program execution.

The authors from \cite{ART:SECONTANNOTRES} propose a security ontology for resource annotation. The proposed ontology defines concepts for security and authorization, for cryptographic algorithms and for credentials. This proposal was designed to be used in the process of security protocol description and selection based on several criteria. In contrast, our ontologies, have a more detailed construction. For example, the ontology from \cite{ART:SECONTANNOTRES} defines a collection of cryptographic algorithms, however, it does not define the algorithm mode, which is a more implementation-specific information. In addition, we did not only propose an ontology, but also a set of rules to construct a specification.

There have been several other security ontologies proposed \cite{ART:ONTCOMP,ART:ONTINFSEC}. Because they do not relate to the specification of security protocols, they can not replace our proposal, but only complete it with additional concepts.

%------------------------------------------------------------------------
\section{Conclusions and future work}\label{SEC:CONCLUSIONS}

We proposed a framework for generating security
protocol specifications. Our proposal generates two components: a message sequential and a semantic component. The first one is implemented through the use of WSDL-S while the second one through OWL.

In order to validate our proposal we constructed over 38
specifications for well-known protocols from the literature (e.g.
SSL, Kerberos) and tested them for automatic execution on client and
server programs.

As future work we intend to create a tool for the automatic
transformation of regular informal specifications into the specifications proposed in this paper based on the described framework.

% We also intend to
%use multiple specifications in the automatic development of composed
%protocols.

%-------------------------------------------------------------------------
%\bibliographystyle{bare_conf.bib}
%\bibliography{bare_conf.bib}

\end{document}